\documentclass{PoS}

\newcommand{\cd}{\! \cdot \!}
\newcommand{\be}{\begin{equation}}
\newcommand{\ee}{\end{equation}}
\newcommand{\ba}{\begin{eqnarray}}
\newcommand{\ea}{\end{eqnarray}}

\title{Shear viscosity in superfluid neutron stars}

\ShortTitle{Shear viscosity in superfluid neutron stars}

\author{\speaker{Laura Tol\'os}\\
     Instituto de Ciencias del Espacio (IEEC/CSIC) Campus Universitat Aut\`onoma de Barcelona, Facultat de Ci\`encies, Torre C5, E-08193 Bellaterra (Barcelona), Spain   \\
        E-mail: \email{tolos@ice.csic.es}}

\author{Cristina Manuel\\
        Instituto de Ciencias del Espacio (IEEC/CSIC) Campus Universitat Aut\`onoma de Barcelona, Facultat de Ci\`encies, Torre C5, E-08193 Bellaterra (Barcelona), Spain}

\abstract{The phonon contribution to the shear viscosity  $\eta$ 
in  superfluid neutron stars is calculated by assuming neutron pairing in a $^1S_0$ channel.
The shear viscosity is obtained by means of variational methods for the solution of the Boltzmann equation amended by a collision term which takes into account the binary collisions of phonons. Effective field theory techniques are used to extract the phonon scattering rates in terms of
the equation of state (EoS) of the system.  We find that $\eta \propto 1/T^5$,
the proportionality factor depending on the EoS of the system.
Our results indicate that the phonon contribution to $\eta$  might have important effects for the different oscillation modes of the star.
}

\FullConference{Sixth International Conference on Quarks and Nuclear Physics,\\
		April 16-20, 2012\\
		Ecole Polytechnique, Palaiseau, Paris}

\begin{document}
\section{Introduction}

The  study of superfluidity in compact stars, such as neutron stars, has received a lot of attention 
after Migdal's observation that superfluidity of neutron matter may occur in the core of compact stars  \cite{Migdal}. In fact, this possibility has been taken into account to explain very different neutron star phenomena, such as pulsar glitches, star cooling or the dynamics of the neutron star's oscillations.

At low temperatures neutron matter superfluidity  occurs after the appearance of a quantum condensate, associated to neutron pairing. The condensate spontaneously breaks the global $U(1)$ symmetry associated to baryon number conservation. Thus, the Goldstone's theorem predicts the existence of a low energy mode which is essential to explain the property of superfluidity, the so-called superfluid phonon.

In this manuscript we will explore the phonon contribution to the shear viscosity  $\eta$  of superfluid neutron matter in the core of the stars, assuming that the neutrons pair in a $^1S_0$ channel \cite{Manuel:2011ed}. In order to compute the phonon contribution  one needs to asses the relevant phonon collisions which are responsible for the transport phenomena. The leading phonon interactions can be completely determined by the equation of state (EoS) of the system \cite{Son:2002zn,Son:2005rv}.  In this paper
we exploit the universal character of the effective field theory (EFT) at leading order to present a very general formulation, that only depends on the equation-of-state (EoS) of the system.  

\section{Phonon interactions in superfluid matter}

The superfluid phonon is the Goldstone mode associated to the spontaneous symmetry breaking
of a $U(1)$ symmetry, which corresponds to particle number conservation. EFT techniques can be used to write down the effective Lagrangian associated to the superfluid phonon. The effective lagrangian is given by an expansion in derivatives of the Goldstone field, the terms of this expansion being restricted by symmetry considerations. The coefficients of the Lagrangian can be computed from the microscopic theory, through a standard matching procedure, thus depending on the EoS of the system.

At the lowest order in a derivative expansion, the Lagrangian reads \cite{Son:2002zn,Son:2005rv}
\begin{eqnarray}
\label{LO-Lagran}
\mathcal{L}_{\rm LO} &=&P (X) \ , \nonumber \\
 X &=& \mu-\partial_t\varphi-\frac{({\bf \nabla}\varphi)^2}{2m} \ ,
\end{eqnarray}
where $P(\mu)$ and $\mu$ are the pressure and chemical potential, respectively, of the superfluid at $T=0$. The quantity
$\varphi$ is the phonon field and $m$ is the mass of the
particles that condense. After a Legendre
transform, one can get the associated Hamiltonian, which has the same form as the one used by Landau to obtain
the self-interactions of the phonons of $^4$He \cite{Son:2005rv,IntroSupe}.

After a Taylor expansion of the pressure, and rescaling of the phonon field to have a canonically normalized kinetic term, the Lagrangian for the phonon field is given by
\begin{eqnarray}
\label{comlag}
\mathcal{L}_{\rm LO}&&=\frac{1}{2}\left((\partial_t\phi)^2-v^2_{\rm ph}({\bf \nabla}\phi)^2\right) \nonumber \\
&&-g\left((\partial_t \phi)^3-3\eta_g \,\partial_t \phi({\bf \nabla}\phi)^2 \right)
+\lambda\left((\partial_t\phi)^4-\eta_{\lambda,1} (\partial_t\phi)^2({\bf \nabla}\phi)^2+\eta_{\lambda, 2}({\bf \nabla}\phi)^4\right)
+ \cdots
\end{eqnarray}
The different  phonon self-couplings of Eq.~(\ref{comlag}) can be  expressed
as different ratios of derivatives of the pressure with respect to the chemical potential \cite{Escobedo:2010uv}, or
in terms of the density, the speed of sound at $T=0$, and derivatives
of the speed of sound with respect to the density. The speed of sound at $T=0$ is given by 
\begin{equation}
\label{phspeed}
v_{\rm ph}=   \sqrt{\frac{\partial P}{\partial {\tilde \rho}} } \equiv c_s \ ,
\end{equation}
where ${\tilde \rho}$ is the mass density, related to the particle density $\rho$ as ${\tilde \rho} = m \rho$.

The  dispersion law obtained from this Lagrangian at tree
level is exactly  $E_p = c_s p $.  And the three and four phonon self-coupling constants can be expressed as 
\ba
g&=&\frac{1}{6 \sqrt{m\rho} \ c_s } \left(1-2 \frac{\rho}{c_s}\frac{\partial c_s}{\partial \rho} \right) \ , \qquad 
\eta_g = \frac{c_s}{6 \sqrt{m\rho}\  g }  \ , \nonumber \\
\lambda &=& \frac{1}{24  \ m  \rho \ c_s^2 } \left( 1-8 \frac{\rho}{c_s} \frac{\partial c_s}{\partial \rho}+10 \frac{\rho^2}{c_s^2} \left( \frac{\partial c_s}{\partial \rho} \right)^2-2\frac{\rho^2}{c_s} \frac{\partial^2 c_s}{\partial \rho ^2}\right)  \ , \nonumber \\
\eta_{\lambda_2} &=&\frac{c_s^2}{8 \ m  \rho \ \lambda } \ , \qquad
\eta_{\lambda,1} = 2 \frac{\eta_{\lambda,2}}{\eta_g}  \ .
\label{eq:relations}
\ea

\section{Equation of state of superfluid neutron matter}
\label{EOS-section}

The speed of sound at $T=0$ as well as the different phonon self-couplings are determined by means of the EoS for neutron matter in neutron stars.  A common benchmark for a nucleonic EoS is the one  by  Akmal,  Pandharipande and  Ravenhall (APR) \cite{ak-pan-rav} in $\beta$-stable nuclear matter. Heiselberg and Hjorth-Jensen  parametrized the APR EoS of nuclear matter in a simple form \cite{hei-hjo}, which will subsequently  be used in this paper. The effect of neutron pairing is not considered  because it is not expected to have a big impact in the EoS. 

Using this EoS, one can obtain the  speed of sound and all three and four phonon self-couplings. The speed of sound can be computed as
\begin{eqnarray}
c_s\approx \sqrt{\frac{1}{m}\frac{\partial P_N}{\partial \rho_n}} , 
\end{eqnarray}
where to compute the mass
density we only take into account the nucleonic part to the pressure ($P_N$),  as $m \gg m_e$, with $m$ and $m_e$ being the mass of the nucleons and electrons, respectively. Further, in  $\beta$-equilibrated matter, $\rho \approx \rho_n$, where $\rho_n$ is the neutron density. For the phonon self-couplings, we will substitute  $\rho \rightarrow \rho_n$ in Eq.~(\ref{eq:relations}). 

\begin{figure}[t]
\begin{center}
\includegraphics[width=0.5\textwidth]{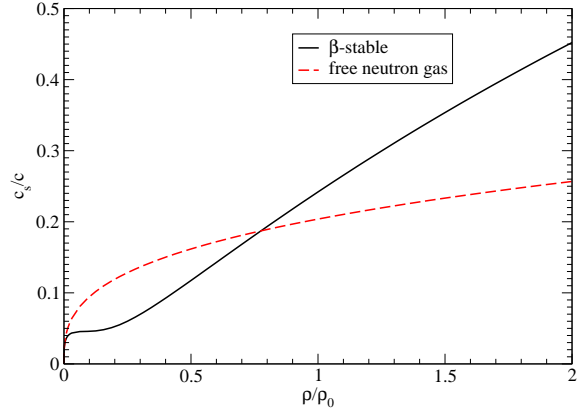}
\caption{ Speed of sound $c_s$ with respect to the speed of light  $c$ for a free gas of neutrons and for $\beta$-stable nuclear matter as a function of the density (taken from Ref.~\cite{Manuel:2011ed})}
\label{speeds}
\end{center}
\end{figure}

The  ratio of the speed of sound $c_s$ with respect to the speed of light $c$ is shown in Fig.~\ref{speeds}  for both a free gas of neutrons and for  $\beta$-stable nuclear matter as a function of the density. We observe that the assumption that the APR model for $\beta$-stable nuclear matter is non-relativistic breaks down at densities of the order of 1.5-2 $\rho_0$. For those densities, relativistic effects appear as the APR EoS includes not only two nucleon but also three nucleon interactions. This seems to suggest that at very high densities we should employ the relativistic version of the phonon effective field theory   \cite{Son:2002zn}. This would imply, though, very minor changes in the computation of $\eta$ (see Ref.~\cite{Manuel:2004iv}).

\section{Shear viscosity due to the binary collisions of  superfluid phonons}

The shear viscosity $\eta$ is as a dissipative term in the energy-momentum tensor $T_{ij}$. For small deviations from equilibrium one
finds that
\be
\label{shear-stress}
 \delta T_{ij}=- \eta \tilde V_{ij}  \equiv  - \eta\left( \partial_i V_j+ \partial_j V_i -\frac 23 \delta_{ij} \nabla \cd {\bf V} \right) \ , 
  \ee
where ${\bf V}$ is the three fluid  velocity of the normal component of the system.

The superfluid phonon contribution to  the energy-momentum
of the system is given by
 \be
 T_{ij}= c_s^2 \int \frac{d^3 p}{(2 \pi)^3}  \frac{ p_i p_j}{E_p} f(p,x)  \ , 
  \ee
where $f$ is the phonon distribution function. The distribution function obeys the Boltzmann equation \cite{IntroSupe}
 \be
  \label{transport}
   \frac{df}{dt} = \frac{\partial
f}{\partial t}+ \frac{\partial E_p}{\partial \bf p} \cdot \nabla f= C[f] \ ,
\ee
being in the superfluid rest frame, and
$C[f]$ is the collision term. For the computation of the collision term it is enough to consider binary collisions. This is due to the fact that the next-to-leading corrections to the dispersion relation of the phonons in superfluid neutron matter implies that the decay of two phonons into two is the first kinematically allowed process \cite{Manuel:2011ed}.

In order to compute the shear viscosity we proceed by considering small departures from equilibrium to the phonon distribution function and linearizing the corresponding transport equation \cite{Manuel:2011ed}. It is then necessary to use variational methods in order to solve the transport equation, as in Refs.~\cite{Manuel:2004iv,Rupak:2007vp,Alford:2009jm}. The final expression for the shear viscosity is given by \cite{Manuel:2011ed}
\be
\eta=\left( \frac{2 \pi}{15} \right)^4 \frac{T^8}{c_s^8} \frac{1}{M} \ ,
\ee  
where $M$ is the is the $2 \leftrightarrow 2$ scattering matrix for phonons. After a very simple dimensional analysis,  we observe that the shear viscosity due to the binary collisions of superfluid phonons scales with the temperature
as  $1/T^5$,  a universal feature that occurs in other superfluid systems  such as ${\rm^4 He}$ \cite{IntroSupe} or superfluid cold atoms at unitary \cite{Rupak:2007vp,Mannarelli:2012su}.

\section{Discussion}

\begin{figure}[t]
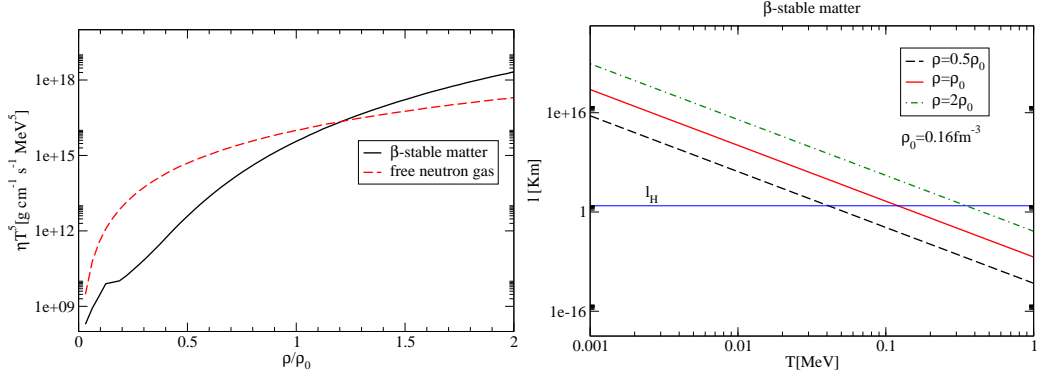

\begin{center}
\includegraphics[width=0.45\textwidth]{shearT5.eps}
\includegraphics[width=0.45\textwidth]{mean_free_path.eps}
\caption{ Left: Shear viscosity multiplied by $T^5$ ($\eta \ T^5$) as function of density for a free neutron gas and $\beta$-stable nuclear matter. Right: Phonon mean-free path in $\beta$-stable matter as function of temperature for different densities. The straight line indicates the hydrodynamic limit: $l_H < 10 \ {\rm Km}$ (taken from Ref.~\cite{Manuel:2011ed}).}
\label{fig:sheart5}
\end{center}
\end{figure}

We start this section by studying the $\rho$ dependence of $\eta$ on the left-hand side of Fig.~\ref{fig:sheart5} as we exploit
that  $\eta T^5 $ is a  temperature-independent quantity. We consider both the APR EoS that describes $\beta$-stable nuclear matter and  the EoS of a free neutron gas.  The viscosity increases by almost three orders of magnitude when the density changes from $\rho_0$ to  2 $\rho_0$ for $\beta$-stable matter, the increase being much more moderate for a free neutron gas. This is related to the behavior of $c_s$ seen in Fig.~\ref{speeds}.   We can then conclude from Fig.~\ref{fig:sheart5}  that the different choices of EoS  have a very clear impact in the numerical values of  $\eta$ while there is a universal $1/T^5$ behaviour with temperature. Needless to say, one should also study other processes such as the phonons colliding with other particles in the star, such as the electrons, in order to assess the dominant processes for dissipation in stars. We leave this computation for future studies.
 
A relevant discussion also is to know the temperature regime  where our results are applicable. Hydrodynamics is only valid when the mean free path is smaller that the typical macroscopic length of the system, in this case, the radius of the star.
We show on the right-hand side of Fig.~\ref{fig:sheart5} the mean free path of phonons within $\beta$-stable matter for $0.5 \rho_0$, $\rho_0$ and $2 \rho_0$ as a function of temperature. We also indicate an estimate of the limit of 10 Km, the radius of the star.  The mean free path $l$ was extracted from the computation of $\eta$ in Ref.~\cite{Alford:2009jm}
\ba
l=\frac{\eta}{n <p>} 
\ea
where $<p>$ is the thermal average momentum, and $n$ is the phonon density:
\ba
<p>=2.7 \frac{T}{c_s} \ , \qquad
n=\int \frac{d^3p}{(2\pi)^3} f_p=\xi(3) \frac{T^3}{\pi^2 c_s^3} .
\ea

On the right-hand side of Fig.~\ref{fig:sheart5}  we observe that for  temperatures below $T\sim 0.1$ MeV, the
phonon mean free path is bigger than the size of the star. This would seem to indicate that
for $T < 0.1$ MeV it is questionable to have a hydrodynamical description of the phonons of the star.
We note that the critical temperature for the phase transition to the normal phase is $T_c \sim 1$ MeV, although
this value, as well as the values we obtained for the phonon mean free path, are dependent on the EoS.

\acknowledgments{ This research was supported by Ministerio de Ciencia e Innovaci\'on under contract FPA2010-16963 and from FP7-PEOPLE-2011-CIG under contract PCIG09-GA-2011-291679. LT acknowledges support from the Ramon y Cajal Research Programme.}

\end{document}